\documentclass[a4paper,fleqn,usenatbib,useAMS]{mnras}
\pdfoutput=1

\usepackage{graphicx}	% Including figure files
\usepackage{amsmath}	% Advanced maths commands
\usepackage{amssymb}	% Extra maths symbols
\usepackage{multicol}        % Multi-column entries in tables
\usepackage{bm}		% Bold maths symbols, including upright Greek
\usepackage{pdflscape}	% Landscape pages
\usepackage{hyperref}

\usepackage{caption}
\usepackage{float}

\newcommand{\eppsilon}{$\varepsilon$ppsilon}
\def\sky{{\theta}}

\usepackage[T1]{fontenc}
\usepackage{ae,aecompl}

\usepackage{txfonts}

\title[EoR Calibration Requirements]{Calibration Requirements for Detecting the 21\,cm Epoch of Reionization Power Spectrum and Implications for the SKA}

\author[N. Barry et. al.]{N. Barry,$^{1}$\thanks{Contact e-mail: \href{mailto:nbarry@uw.edu}{nbarry@uw.edu}} B. Hazelton,$^{2,1}$ I. Sullivan,$^{3}$ M. F. Morales,$^{1}$ J. C. Pober $^{4}$
\\ 
$^{1}$University of Washington, Department of Physics\\
$^{2}$University of Washington, eScience Institute\\
$^{3}$University of Washington, Department of Astronomy\\
$^{4}$Brown University, Department of Physics
}

\date{}

\pubyear{2016}

\begin{document}
\label{firstpage}
\pagerange{\pageref{firstpage}--\pageref{lastpage}}
\maketitle

\begin{abstract}

21\,cm Epoch of Reionization observations promise to transform our understanding of galaxy formation, but these observations are impossible without unprecedented levels of instrument calibration. We present end-to-end simulations of a full EoR power spectrum analysis including all of the major components of a real data processing pipeline: models of astrophysical foregrounds and EoR signal, frequency-dependent instrument effects, sky-based antenna calibration, and the full PS analysis. This study reveals that traditional sky-based per-frequency antenna calibration can only be implemented in EoR measurement analyses if the calibration model is unrealistically accurate. For reasonable levels of catalogue completeness, the calibration introduces contamination in otherwise foreground-free power spectrum modes, precluding a PS measurement. We explore the origin of this contamination and potential mitigation techniques. We show that there is a strong joint constraint on the precision of the calibration catalogue and the inherent spectral smoothness of antennae, and that this has significant implications for the instrumental design of the SKA and other future EoR observatories.

\end{abstract}


%%%%%%%%%%%%%%%%%%%%%%%%%%%%%%%%%%%%%%%%%%%%%%%%%%

%%%%%%%%%%%%%%%%% BODY OF PAPER %%%%%%%%%%%%%%%%%%

%%%%%%%%%%%%%%%%%%%%%%%%%%%
\section{Introduction}
\label{sec:intro}

Observations of the Epoch of Reionization (EoR) promise to reveal a wealth of information about the dynamics and evolution of the universe. The 21\,cm hyperfine transition line of neutral hydrogen is one of the best probes of the EoR (see \citet*{furlanetto_cosmology_2006,morales_reionization_2010} for reviews) and several experiments are currently seeking or will seek a power spectrum (PS) measurement of this faint cosmological signal. The low-frequency interferometers attempting to make these measurements include the Donald C. Backer Precision Array for Probing the Epoch of Reionization (PAPER; \citealt{parsons_precision_2010})\footnote{\url{http://eor.berkeley.edu}}, the LOw Frequency ARray (LOFAR; \citealt{yatawatta_initial_2013,van_haarlem_lofar:_2013})\footnote{\url{http://www.lofar.org}}, the Murchison Widefield Array (MWA; \citealt{lonsdale_murchison_2009,tingay_murchison_2013,bowman_science_2013})\footnote{\url{http://www.mwatelescope.org}}, the Hydrogen Epoch of Reionization Array (HERA; \citealt{pober_what_2014})\footnote{\url{http://reionization.org}}, and the Square Kilometre Array (SKA; \citealt{mellema_reionization_2013,koopmans_cosmic_2015})\footnote{\url{https://www.skatelescope.org}}. 

However, astrophysical foregrounds are 4--5 orders of magnitude brighter than the expected cosmological signal. Suppressing this foreground contamination in the PS requires unprecedented precision in instrumental calibration. We simulate sky-based calibration on the EoR PS signal to explore the techniques necessary to suppress the foreground contamination. Our end-to-end simulations include chromatic instrumental effects common to current EoR experiments and realistic differences between the true sky and the calibration catalogue. We show traditional per-frequency antenna calibration techniques contaminate the EoR PS signal, demonstrate how to improve the calibration procedure for EoR measurements, and place joint constraints on the spectral smoothness of the antennas and the precision of the calibration catalogues needed to reveal the EoR.

Variance and convergence statistics of calibration in image space have been studied in detail, including the application of ionospheric changes \citep{van_der_tol_self-calibration_2007, mitchell_real-time_2008, wijnholds_van_der_veen_multisource_2009, datta_detection_2009}, variation of diffuse structure scales \citep{mitchell_real-time_2008}, addition of polarized components \citep{mitchell_real-time_2008}, inclusion of small source position offsets \citep{ng_sensor-array_1996, wijnholds_van_der_veen_multisource_2009, datta_detection_2009}, and the imperfection of source models \citep{datta_detection_2009, datta_bright_2010}. This includes the approach that we label as traditional, where calibration solutions are solved for each antenna and each frequency from a least squares analysis of the visibilities \citep{salvini_fast_2014}. Calibration techniques and effects have been rigorously tested via variance statistics. However, only a few studies have investigated calibration effects on the PS (e.g. \citealt{trott_2016,switzer_interpreting_2015}). The EoR measurement is to occur in PS space, so the standard of the effectiveness for calibration should also be established in PS space. 

In addition, it is imperative to use a realistically imperfect calibration catalogue for sky-based calibration when simulating resultant effects. Models of sources generated from knowledge of the sky are used to calibrate the instrument \citep{mitchell_real-time_2008}, but no catalogue is perfect. There will always be unmodelled faint sources, either due to confusion limits or the inability to resolve morphology, and catalogues will always include small errors in flux, position, or compactness of sources. These imperfections in the sky model will affect calibration, and subsequently the EoR PS.

This study explores the impact of calibrating against an incomplete source model on the EoR PS measurement. We show that unmodelled, faint sources can interact with calibration to mix foregrounds into unrelated modes. Spectral structure due to the incomplete source model propagated via traditional per-frequency antenna calibration couples bright foreground power from unimportant Fourier modes into the most sensitive EoR modes. Consequently, the EoR measurement is impossible without the development of new calibration techniques beyond traditional methods. 

Descriptions of the simulations, software packages, catalogue data, and PS space are given in \S\ref{sec:methods}. The effects on the PS of using traditional per-frequency calibration techniques are shown in \S\ref{sec:trad}. Mitigation techniques to avoid fitting spectral structure from faint sources are demonstrated in \S\ref{sec:smooth_full}, which also highlights the importance of a spectrally smooth instrumental response. Approaches to faithfully reconstruct true instrumental spectral structure, while minimizing the effect of unmodelled faint sources, are described in \S\ref{sec:bp} and the implications for the SKA and other future EoR machines are discussed in \S\ref{sec:conclusions}. 

We note that this work concentrates on the ``imaging'' EoR PS analysis approach, and the calibration requirements for ``delay'' PS analyses with redundant arrays may be different \citep{parsons_sensitivity_2012}. A full study of the calibration requirements of delay spectra is left for future work.

%%%%%%%%%%%%%%%%%%%%%%%%%%%

%%%%%%%%%%%%%%%%%%%%%%%%%%%
\section{Methods and measurement space}
\label{sec:methods}

Our calibration simulations employ a suite of packages designed for MWA EoR analysis. These full end-to-end simulations demonstrate the effect of calibration errors in the two-dimensional power spectrum (2D PS) --- a primary figure of merit for the EoR measurement. In this section, we describe the 2D EoR PS figure of merit and our simulation methods. 

\subsection{The 2D power spectrum}
\label{sec:PS}

The 21\,cm hyperfine transition of neutral hydrogen is a narrow emission line. This allows the measured frequency of the emission to map closely to its line-of-sight distance. Measurements of the EoR are inherently three dimensional, with two angular dimensions and one frequency dimension represented in the volume $\{\sky_{x}, \sky_{y}, \nu\}$. Using the angular diameter and line-of-sight distances, the observations can be mapped to cosmological coordinates $\{x,y,z\}$ in co-moving Mpc \citep{hogg_distance_1999}. Statistical measurements of the EoR show the most promise for a robust detection in wavenumber space \citep{morales_toward_2004}, represented by $\{k_{x}, k_{y}, k_{z}\}$ and accessible through Fourier transforms. 

\begin{figure}
\centering
	\includegraphics[width = \columnwidth]{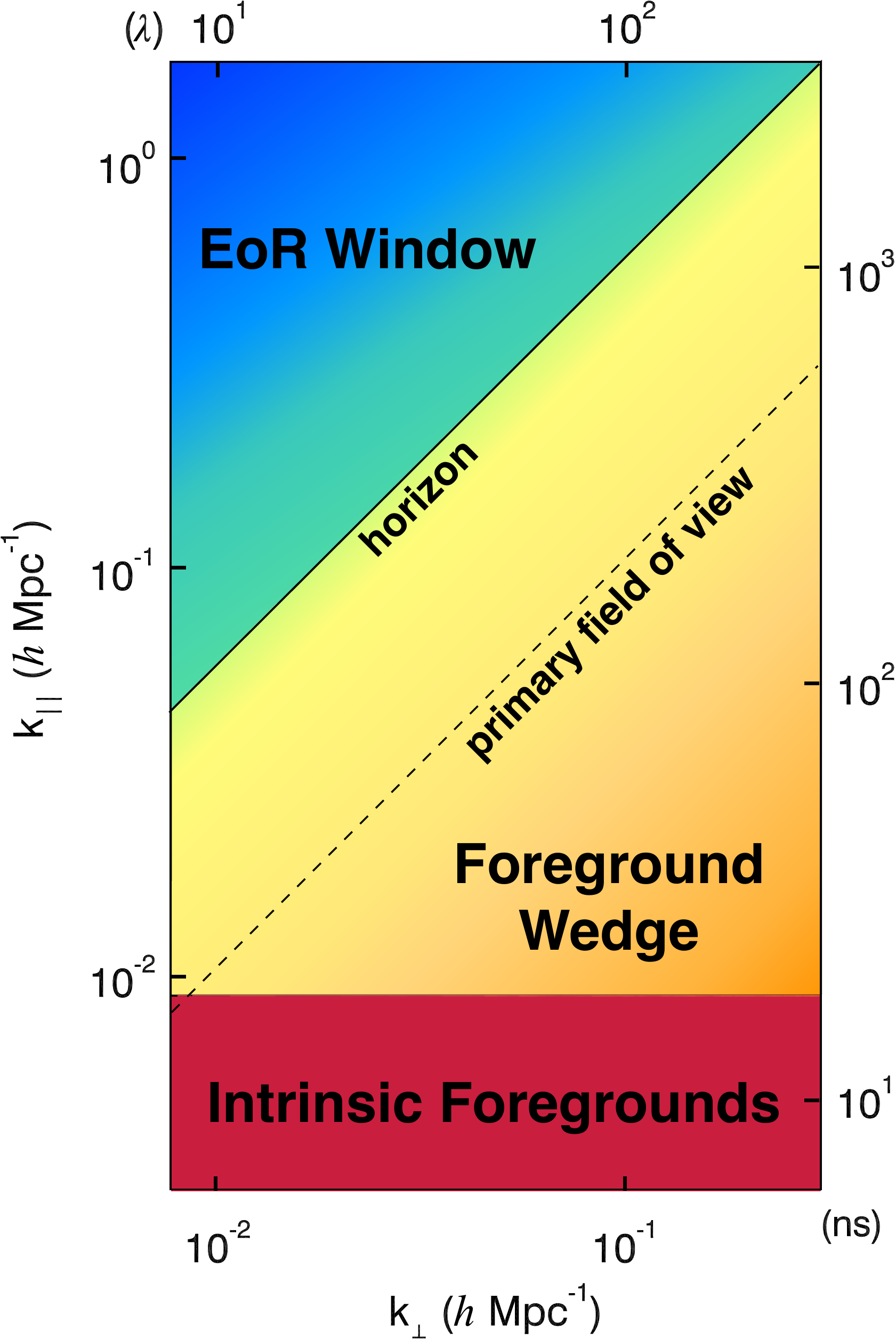}
	\caption{A schematic representation of an expected 2D PS. ``Intrinsic foregrounds'' dominate low $k_{\parallel}$ (modes along the line-of-sight) for all $k_{\bot}$ (modes perpendicular to the line-of-sight) due to their relatively smooth spectral structure. Chromaticity of the instrument mixes foreground modes up into the  ``foreground wedge''. The ``primary field of view'' line and the ``horizon'' line are contamination limits dependent on how far off-axis sources are on the sky. The most sensitive, foreground-free measurement modes are expected to be in the lower, left-hand corner of the ``EoR window.''}
	\label{fig:cartoon}
\end{figure}

The distribution of hydrogen in the universe is isotropic and homogeneous to first order. This spherical symmetry can be harnessed to achieve greater sensitivity. Averaging the squared measurements in spherical shells within the volume $\{k_{x}, k_{y}, k_{z}\}$ allows a transformation into a one-dimensional power spectrum (1D PS). While this aids in the measurement of the EoR, it removes the ability to view the k-space distributions of foreground and calibration effects. We will therefore present results in the 2D PS, achieved through averaging squared measurements along only the angular wavenumbers $\{k_{x},k_{y}\}$. This creates the PS as a function of modes perpendicular to the line-of-sight ($k_\bot$) and modes parallel to the line-of-sight ($k_\parallel$) as shown in Figure~\ref{fig:cartoon}.  Axes are displayed in units of Hubble constant (\emph{h}) times inverse megaparsec (Mpc$^{-1}$).

Wavenumber space is crucial for statistical measurements due to the spectral characteristics of the foregrounds. Diffuse synchrotron emission and bright radio sources, while distributed across the sky, vary smoothly in frequency (e.g.\ \citealt{matteo_radio_2002,peng_oh_foregrounds_2003}). Only small $k_\parallel$ values are theoretically contaminated by bright, spectrally smooth astrophysical foregrounds. Since the foreground power is restricted to only a few low $k_\parallel$ modes, larger $k_\parallel$ values tend to be free of ``intrinsic foregrounds'' in wavenumber space.

However, interferometers are naturally chromatic. This chromaticity distributes foreground power into a distinctive ``foreground wedge'' due to the mode-mixing of power from small $k_\parallel$ values into larger $k_\parallel$ values as demonstrated in Figure~\ref{fig:cartoon} \citep{datta_bright_2010,morales_four_2012,vedantham_imaging_2012,parsons_per-baseline_2012,trott_impact_2012,hazelton_fundamental_2013,thyagarajan_study_2013,pober_opening_2013,liu_epoch_2014}. The ``primary field of view'' line and the ``horizon'' line are the expected contamination limits caused by measured sources in the primary field of view and the sidelobes, respectively. The remaining region, called the ``EoR window,'' is expected to be contaminant-free. Because the power of the EoR signal decreases with increasing $\lvert k \rvert$, the most sensitive measurements are expected to be in the lower, left-hand corner of the EoR window.

\begin{figure*}
\centering
	\includegraphics[width=\textwidth]{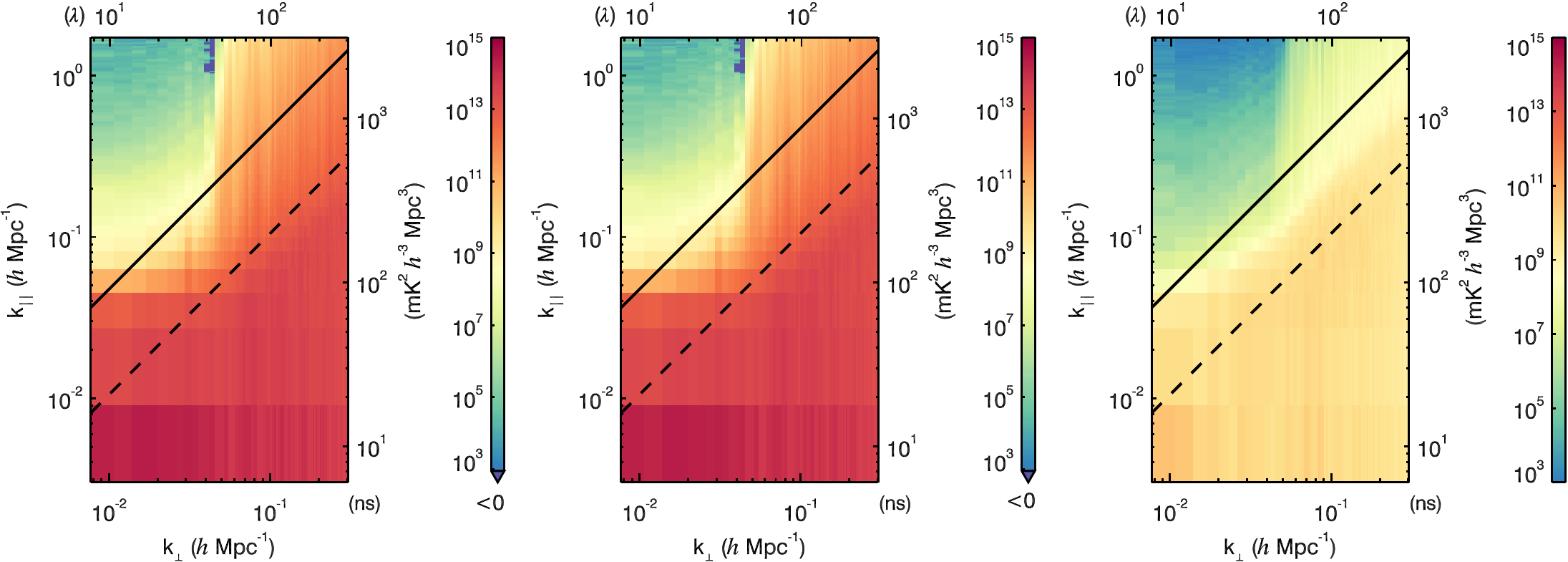}
	\caption{From left to right: the calibrated data from a calibration simulation, the model given known foregrounds and instrumental effects, and the residual after the model is subtracted from the calibrated data. The residual 2D PS has the potential to reveal more modes to the EoR PS measurement given the accuracy and completeness of the model.}
	\label{fig:panel}
\end{figure*}

PS from our end-to-end simulation in Figure~\ref{fig:panel} show the standard features described in the Figure~\ref{fig:cartoon} schematic. The left plot is the 2D PS of the calibrated simulation containing foregrounds and EoR signal. The middle plot, which looks nearly identical, shows the instrument calibration model. This contains a subset of the foregrounds to simulate an incomplete knowledge of the sky. We can decrease the contamination in the PS by subtracting this model from calibrated data to possibly reveal the EoR signal in a wider range of modes, depending on completeness of the model. Taking the difference yields the ``residual'' 2D PS in the right plot of Figure~\ref{fig:panel}, which reveals unmodelled foregrounds with their instrumental effects and the EoR signal. This subtraction is implemented in the three dimensional measurement cube before constructing the PS. In addition to the typical 2D PS effects, foregrounds also contaminate higher k$_\bot$ as a consequence of decreased baseline coverage of the instrument at those scales, which is specific and intrinsic to each array. For the remainder of this paper, we will use the residual 2D PS space to explore the effects of calibration.

\subsection{Simulation methods}
\label{sec:sim}

All of the calibration simulations in this paper utilize the MWA antennae and positions in the frequency band 167--198\,MHz. We use a precursor to the KGS catalogue (Carroll et al.\ in review) as our foreground model and the Fast Holographic Deconvolution\footnote{FHD software package is available at \url{https://github.com/EoRImaging/FHD}} (FHD) software package to implement our simulations, calibration, and imaging \citep{sullivan_fast_2012}. To create PS, we use the Error Propagated Power Spectrum with InterLeaved Observed Noise\footnote{\eppsilon{} software package is available at \url{https://github.com/EoRImaging/eppsilon}} (\eppsilon) software package. This simulates the full end-to-end MWA PS analysis detailed in \citet{jacobs_murchison_2016}. 

A model of 6950 compact sources seen by the MWA and compiled in the KGS catalogue were used as simulated input data, along with the addition of a simulated Gaussian EoR signal in the visibilities. This approach is completely noiseless, and contains no information about ionospheric effects or diffuse galactic emission. 

In addition to the simulated input data, a model of the sky is generated for sky-based calibration. Antenna gain solutions that minimize the differences in visibilities between the input data and the calibration model are calculated through an iterative, least-squares approach using all cross-correlated visibilities \citep{salvini_fast_2014}. The final result constitutes our calibration solutions, which are used to generate snapshot PS for one observation.

\begin{figure}
	\includegraphics[width = \columnwidth]{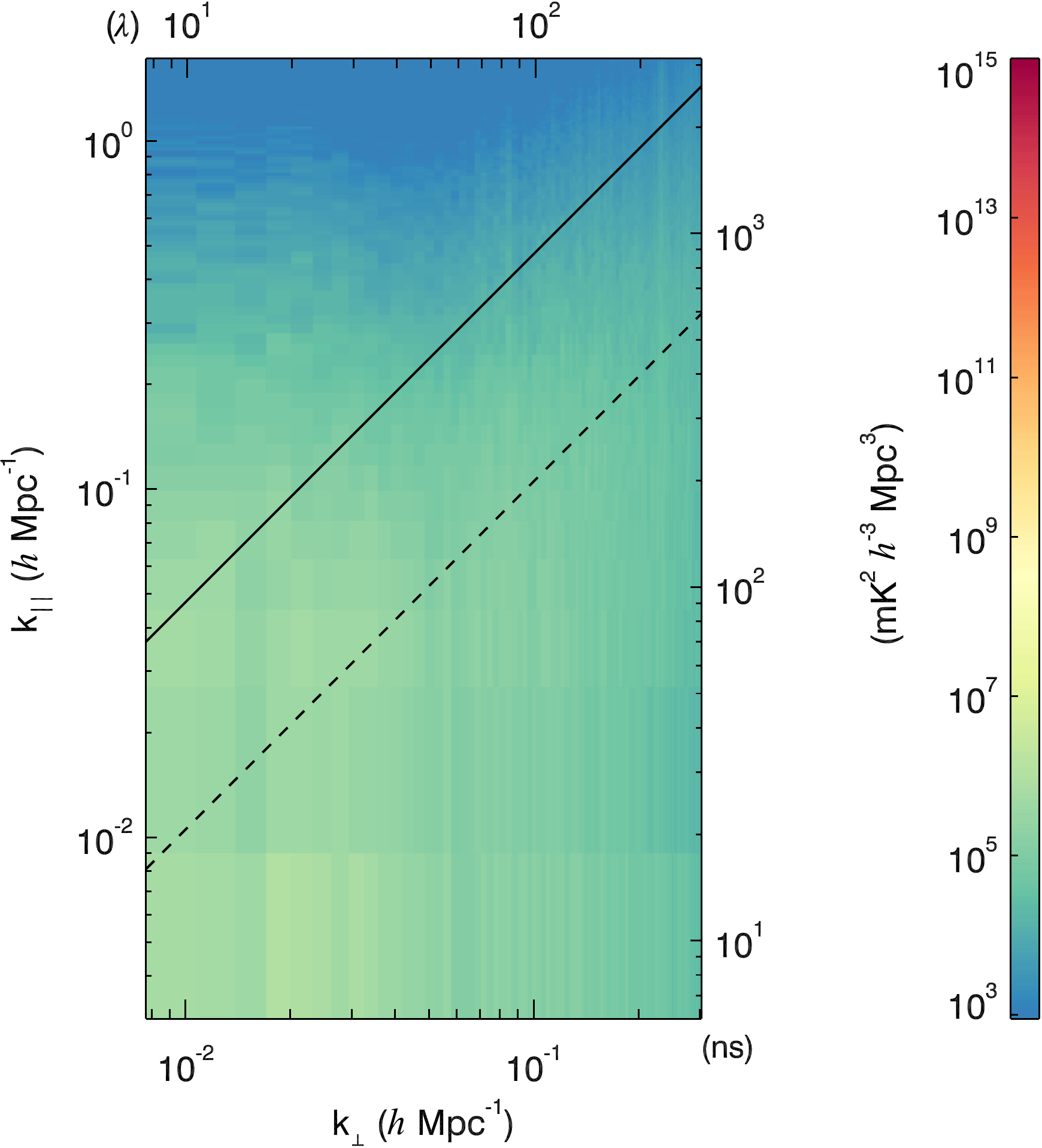}
	\caption{The result of the calibration simulation pipeline with a perfect sky model. Modelling, calibrating, and subtracting all of the 6950 KGS sources using FHD and \eppsilon{} recovers the added simulated EoR signal. The EoR signal that we recover does not experience signal loss regardless of calibration technique used. The color scale has been fixed throughout this paper to provide order of magnitude reference.}
	\label{fig:eor}
\end{figure}

When we allow the simulation to use all of the input catalogue sources as a model from which to calibrate and subtract, all foreground sources are removed perfectly. This reveals the simulated EoR signal in the residual PS with no foregrounds or chromaticity effects, as seen in Figure~\ref{fig:eor}. The color scale has been fixed to highlight the order of magnitude difference between the EoR signal peaked at $10^{6}$\,mK$^2$\,\emph{h}$^{-3}$\,Mpc$^3$ and the intrinsic and mode-mixed foregrounds peaked at $10^{14}$\,mK$^2$\,\emph{h}$^{-3}$\,Mpc$^3$ in Figures~\ref{fig:cartoon}~\&~\ref{fig:panel}. Retrieving the EoR PS demonstrates consistency within the simulation, and will provide a magnitude scale for simulation outputs with unmodelled, faint sources. This also demonstrates that the pipeline effectively recovers the EoR signal with very little contamination or signal loss if the foreground model is perfect, regardless of specific calibration techniques.

To test the effects of an incomplete source model in the analysis pipeline, we then build a new catalogue with only the 4000 brightest sources as seen by the instrument from our KGS catalogue. The new catalogue is used for both calibration and subtraction. This left 2950 of the faintest sources unmodelled and unsubtracted, with an average flux of 0.7\,Jy and minimum flux of 0.15\,Jy. These faintest sources account for 39\% of the total flux. In addition, only 16\% of them are actually in the field of view --- the rest are in antenna sidelobes.  While catalogues are becoming more precise and the errors are decreasing in both the field of view and sidelobes \citep{pober_importance_2016}, this level of accuracy illustrates the types of errors that will arise in any measurement with an imperfect calibration sky catalogue. This provides an investigation into the contribution of an incomplete source model to the EoR PS measurement through calibration.

All effects resulting from our calibration simulations can be purely attributable to which sources from the input catalogue are calibrated and subtracted. There are \textit{no position, flux, morphology, or beam errors}, as well as \textit{no direction-dependent, ionospheric, diffuse emission, or polarization effects.} These other realistic factors are dealt with in various ways after an initial source catalogue calibration. 

We explore the spectral contamination associated with this initial source catalogue calibration step. However, additional calibration techniques that account for the other realistic effects will also contribute to spectral contamination. Our focus on source catalogue calibration therefore provides a base level of expected contamination. Iterative calibration techniques such as self-calibration \citep{pearson_image_1984} will also have very similar errors due to the mismatch between the final model and true sky, and can be simulated using the same technique used here.

%Many calibration pipelines use additional models and direction-dependent techniques after an initial calibration on a source model (cite LOFAR, MWA, any others?). Our simulations explore the effects of a source catalogue calibration step, which illustrates a result that all pipelines can experience if source catalogue calibration is required. Additional errors and contamination from other calibration steps to mitigate other realistic 

%move realistic effects into italics. stronger wording on the fact that source cal is first step, and any other cal done will only add error.

%%%%%%%%%%%%%%%%%%%%%%%%%%%

%%%%%%%%%%%%%%%%%%%%%%%%%%%
\section{Calibration errors due to faint, unmodelled  sources}
\label{sec:trad}

Traditional radio astronomy calibration techniques involve calculating the gain at every frequency for every antenna from the visibilities using an iterative least squares solver. Historically, this approach was an extension from single-frequency radio astronomy to small-bandwidth multi-frequency instrumentation \citep{fomalont_1999,iii_cosmic_2009}.  While this method involves solving for many variables, the number of degrees of freedom in the data from the MWA is orders of magnitude larger than the number of parameters used and thus theoretically constrained.  This calibration method has remained a stalwart in the community as the field has advanced. Our calibration simulations first examine the traditional radio astronomy calibration technique and how it effects the EoR PS measurement.

\begin{figure*}
\centering
	\includegraphics[width=\textwidth]{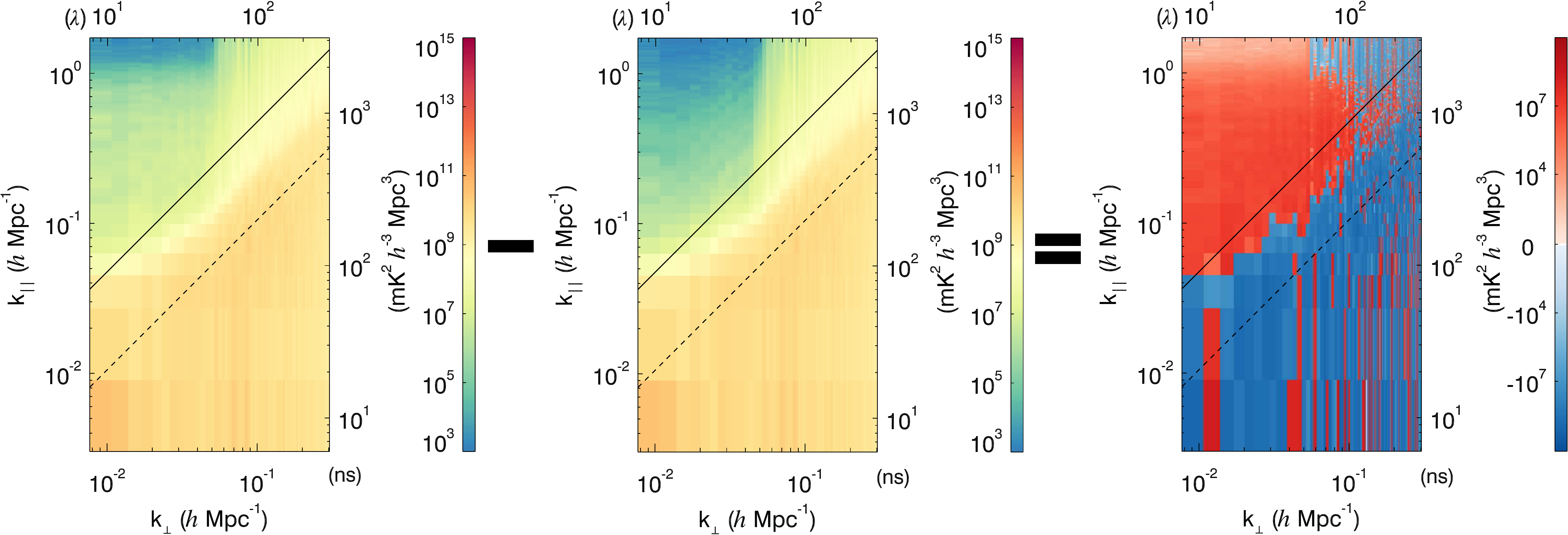}
	\caption{The subtraction of a residual 2D PS with traditional per-frequency antenna calibration methods (left) and a reference residual 2D PS without simulated calibration effects (middle) to create a difference 2D PS (right). Red indicates a relative excess of power, and blue indicates a relative depression of power. Spectral contamination power at all modes in the EoR window is evident. The most sensitive, theoretically contaminant-free EoR modes have excess power on levels of $10^7\,$mK$^2\,$\emph{h}$^{-3}\,$Mpc$^3$, making the measurement impossible with reasonable calibration catalogue errors and traditional per-frequency antenna calibration.}
	\label{fig:trad_diff}
\end{figure*}

To capture realistic differences between the true sky and the calibration catalogue, we simulate the sky as 6950 sources but only use the brightest 4000 to predict the visibilities for use in calibration. This introduces small differences between the sky and calibration visibilities that can affect the per-frequency antenna calibration solutions.
We apply the antenna calibration solutions to the input sky visibilities, and then subtract the brightest 4000 sources used in the calibration model for one observation. This residual 2D PS is shown as the left plot of Figure~\ref{fig:trad_diff}. The 2950 unmodelled faint sources populate the foreground wedge as expected. 

We can also calculate a residual 2D PS with a perfect calibration and with the same 4000 source foreground subtraction to provide a reference for the observation. This is shown in the middle panel of Figure~\ref{fig:trad_diff} (which is the same residual 2D PS shown in the righthand panel of Figure~\ref{fig:panel}). Unmodelled faint sources also populate the foreground wedge; however, it is apparent that the 2D PS using traditional per-frequency antenna calibration has relatively high amounts of power in the EoR window.

The calibration simulation can be used to quantify the shape and amount of excess power in PS space. Direct subtraction between the traditional calibration simulation 2D PS (the left plot of Figure~\ref{fig:trad_diff}) and the reference 2D PS (the middle plot of Figure~\ref{fig:trad_diff}) provides this information. The result is a difference 2D PS, where red indicates relative excess power and blue indicates relative depressed power. Evident in the difference 2D PS is the excess power contamination of the entire EoR window by as much as $10^7\,$mK$^2\,$\emph{h}$^{-3}\,$Mpc$^3$. It is important to note that this level of calibration error would make the EoR measurement impossible. Using traditional per-frequency antenna calibration in a PS measurement would require a highly accurate calibration catalogue. 

Qualitatively, allowing independent calibration parameters for each frequency channel and antenna allows small deviations from the true solutions on small spectral scales. These amplitude and phase deviations are caused by the point spread functions (PSF)  of unmodelled sources which modify the observed fluxes of true sources, as seen in real data by \citet{offringa_parametrising_2016}. This effect is frequency dependent and its magnitude depends on the completeness of the sky model and the natural PSF of the array. The resulting calibration errors are only on the order of 1 part in 10$^3$ in this simulation. However, this varied spectral structure in the calibration solutions is enough to couple power from the bright, intrinsic foregrounds to the Fourier modes in the EoR window. This fills every possible EoR measurement mode with foreground power. 

Not only are sensitive regions of the EoR window dominated by coupled power from intrinsic foregrounds, but there is a corresponding depression of power in the foreground wedge as well. This is also the result of small spectral deviations captured in the calibration solutions. The measured fluxes of modelled sources do not accurately reflect the true fluxes due to the residual PSF of unmodelled sources. Allowing calibration solutions to be modified by this residual structure results in overfitting and over-subtraction.

Using the modulation theorem, we can quantitatively associate the level of contamination seen in the PS with the observed calibration errors. Data that is modified by spectrally variant calibration solutions is Fourier transformed into PS space, and the modulation theorem of Fourier transforms results in mode-mixing between the modes of the unmodelled spectral structure and the bright foreground wedge. This couples the response of foregrounds with calibration deviations along the frequency axis. 

Excess power can be estimated given a modulated signal 
\begin{equation}
h(\nu)=f(\nu)\left(1+\Delta g\cos{\eta_0\nu}\right),
\end{equation}
where $h(\nu)$ is the modulated instrumental response as a function of frequency, $f(\nu)$ is the original instrumental response as a function of frequency, $\eta_0$ is the Fourier dual of a mode in the amplitude deviations of the calibration gain, and $\Delta g$ is the amplitude deviation associated with the frequency mode $\eta_0$. The modulation theorem results in the Fourier transform
\begin{equation}
H(\eta)=\frac{\Delta g}{2}F(\eta-\eta_0)+\frac{\Delta g}{2}F(\eta+\eta_0)+ F(\eta).
\label{eq:FT}
\end{equation}
Fourier transforms of the original signal $f$ constructs signal at $\eta$, $\eta-\eta_0$, and $\eta+\eta_0$. Equation~\ref{eq:FT} is squared to obtain the PS, and cross-terms between $F(\eta)$ and $F(\eta\pm\eta_0)$ can be neglected since overlap is small for an $\eta_0$ in the EoR window. An order of magnitude estimate of the positive power spectrum of this modified signal is 
\begin{equation}
\mathcal{O}(\lvert H(\eta)\rvert^2) \approx \mathcal{O}(\lvert F(\eta)\rvert^2) + \mathcal{O}\left(\left| \frac{\Delta g}{2} F(\eta\pm\eta_0)\right|^2\right).
\end{equation}
As a result, the modulated power response $\mathcal{O}(\lvert H(\eta)\rvert^2)$ has power contributions as a function of $\eta$ and, to a lesser extent, $\eta\pm\eta_0$. When all $\eta$ and $\eta_0$ values are considered, the result is equivalent to the convolution of the foregrounds with the Fourier transform of the calibration deviations.

For small $\eta$ values, intrinsic foregrounds dominate. Power will be modulated from these intrinsic foregrounds into any frequency mode $\eta_0$ captured in the amplitude deviations in calibration. Given simulation values of the intrinsic foregrounds ($\mathcal{O}(P_{k_0}) \approx 10^{14}$\,mK$^2$\,\emph{h}$^{-3}$\,Mpc$^3$) and the amplitude deviations ($\mathcal{O}(\lvert \frac{\Delta g}{2 } \rvert^2) \approx  10^{-7}$, or a $\Delta g$ of order 1 part in 10$^3$), the excess contamination in frequency mode $\eta_0$ of the PS is estimated to be $10^{7}$\,mK$^2$\,\emph{h}$^{-3}$\,Mpc$^3$. This agrees with the level of contaminated power in Figure~\ref{fig:trad_diff} generated by calibration simulations.

The satisfactory performance of traditional per-frequency antenna calibration depends on a highly accurate calibration catalogue. When we use the same sources to generate the sky and calibration models --- even with an added EoR signal --- the resulting calibration and foreground suppression in the PS is excellent, as seen in Figure~\ref{fig:eor}. However, this is not a realistic situation for current and planned EoR observatories. When the calibration catalogue is not perfect, traditional per-frequency antenna calibration distributes spectral power and overwhelms the faint cosmological signal as seen in Figure~\ref{fig:trad_diff}. This sets very strong constraints on the accuracy of the calibration catalogue if the traditional calibration approach is to be used for EoR measurements.

%%%%%%%%%%%%%%%%%%%%%%%%%%%

%%%%%%%%%%%%%%%%%%%%%%%%%%%
\section{Mitigation by smooth calibration solutions}
\label{sec:smooth_full}

Spectral contamination in the EoR window from traditional calibration techniques necessitates mitigation. If the instrument is spectrally-smooth across the frequency band, we can use this as a prior that must be met in our calibration solutions. We explore constraining the spectral variation of the calibration to be smooth relative to the band size to avoid contamination of the EoR window. However, non-smooth spectral features of the instrument must be incorporated into the calibration, and therefore we also investigate the consequences of fitting specific instrumental features.

\subsection{Constraining smooth instrumental response}
\label{sec:smooth}

If an antenna has a naturally smooth bandpass, its response can be modeled with low-order polynomials or other slowly varying functions. With this restriction, we avoid the fine-scale spectral structure in the calibration solutions that causes the contamination of the EoR window seen in Section~\ref{sec:trad}. Perfect calibration solutions are flat in our simulation, so polynomials applied to the frequency band would only model the level of error expected with polynomial fitting.

We calculate best-fit polynomials over the whole frequency band from the traditional calibration solutions generated in Section~\ref{sec:trad} with a calibration catalogue of the brightest 4000 of the 6950 input sources. Five calibration parameters for the frequency band are allowed and are chosen to represent typical instrumental variation. Three amplitude parameters create a second-order bandpass-like polynomial, and two phase parameters create a smooth ramp in frequency.

\begin{figure}
	\includegraphics[width = \columnwidth]{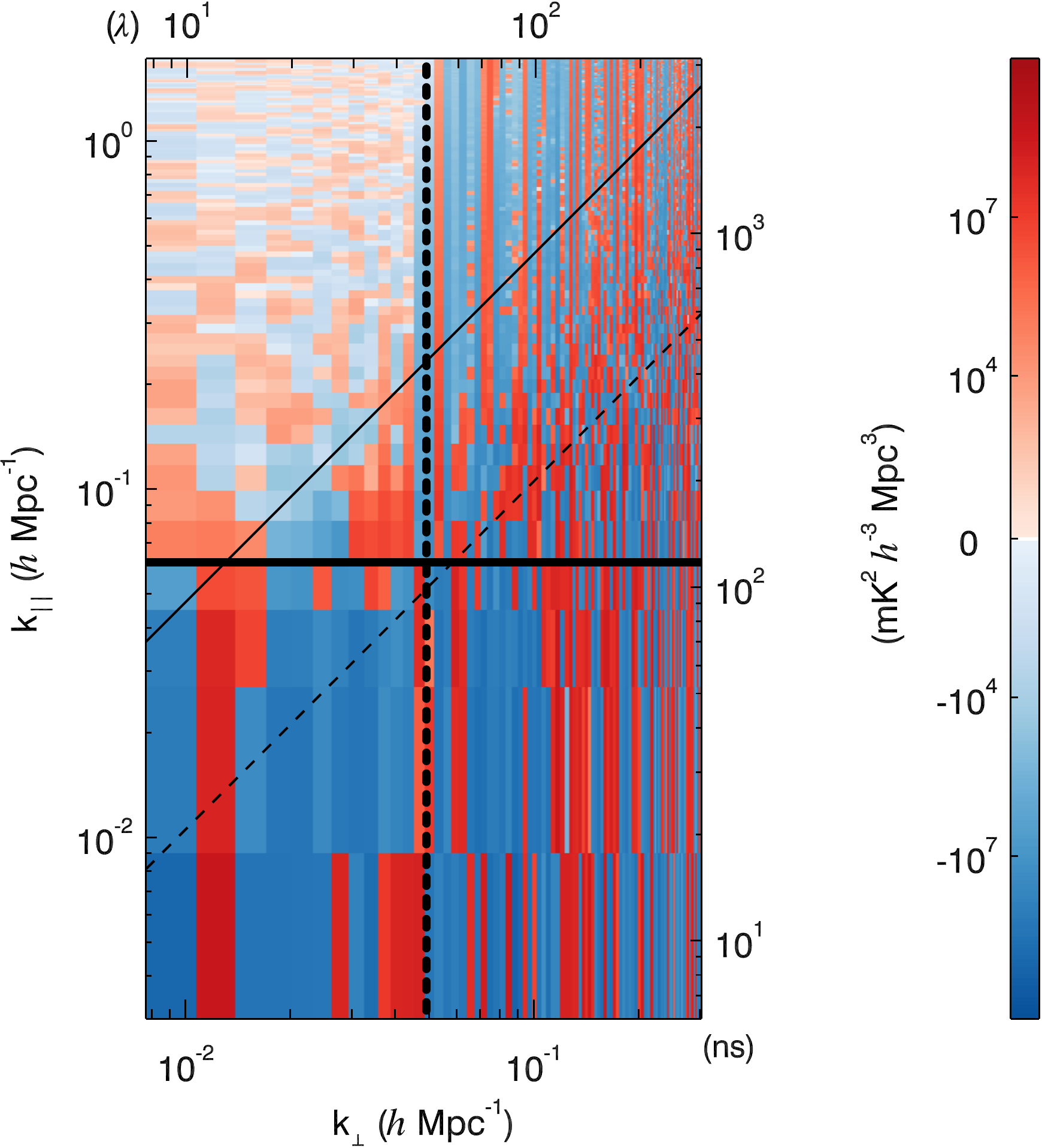}
	\caption{The difference 2D PS between a reference 2D PS and a 2D PS where calibration solutions were modelled with a second-order polynomial in amplitude and a linear fit in phase. Red indicates an excess of power, and blue indicates a depression of power. The bold line indicates the largest k$_\parallel$ affected by power spectral contamination from low-order polynomial fitting. The dotted line indicates the largest k$_\bot$ not affected by the coupling of poor baseline coverage to the foreground wedge. The EoR window is noise-like and unbiased since smooth mitigation techniques did not capture fine-scale spectral structure caused by unmodelled sources.}
	\label{fig:smooth}
\end{figure}

Figure~\ref{fig:smooth} shows the difference 2D PS between the smooth calibration solution PS and the reference PS for one observation. The EoR window from the smooth mitigation technique and the reference are strikingly similar, leading to very little difference. The level of the difference is also noise-like and far below the EoR signal. This will neither affect an EoR measurement to a significant degree nor bias the result.

Power from intrinsic foregrounds is not coupled to the EoR window due to the restriction of smoothness relative to EoR spectral modes. Spectral contamination on the scale of the polynomials still occurs; however, this contamination only occurs within the foreground wedge and will not hinder the EoR measurement. The bold line in Figure~\ref{fig:smooth} highlights the highest k$_\parallel$ with significant contamination caused by the low-order polynomial fitting, which is below the EoR window.

Differences in power in the region of high k$_\bot$ and high k$_\parallel$ in Figure~\ref{fig:smooth} are also apparent. Poor baseline coverage couples the foreground wedge to this region as described in Section~\ref{sec:PS}. Since spectral contamination did occur in the foreground wedge, power changes occur in the region affected by poor baseline coverage. EoR measurements will not be made in k-space areas with poor baseline coverage, so power changes in this area are not a large concern. The dotted line in Figure~\ref{fig:smooth} indicates the largest k$_\bot$ with high baseline coverage. A significant foreground-free EoR window to the upper left remains.

By restricting the instrumental response to be smooth with respect to EoR spectral modes, we significantly reduced the excess power in the EoR window caused by spectral contamination. Now, the EoR window has no power bias and what little contamination there is appears noise-like. Measuring the EoR signal in a 2D PS that utilizes the smooth mitigation technique in the calibration solutions will be essentially unaffected by an imperfect catalogue. Instrumentation that is spectrally smooth can avoid contaminating the EoR window using this technique.

%%%%%%%%%
\subsection{Calibration parameters in spectral modes}	
\label{sec:150m}

Instrumental responses are not always smooth across the frequency band. Any spectral features in the instrument need to be fit so that the calibration is physically true. We simulate the effect of calibrating cable reflections as an example of the consequences of fitting for instrumental structure on a per-antenna basis.

Receiver-to-beamformer cable reflections with amplitudes of $\sim$1\% of the signal are apparent in MWA data, creating a characteristic frequency ripple in the antenna gain at the corresponding light travel time \citep{dillon_empirical_2015,beardsley_thesis,ewall-wice_2016_xray,jacobs_murchison_2016}. Cable reflections can vary dramatically from antenna to antenna, and must be fit individually. Our calibration simulation uses three parameters (mode, amplitude, and phase) to describe the spectral ripple from a hypothetical 150\,m cable reflection in a subset of the antennas. The 150\,m cable reflection does not actually exist in the simulated data, thus the fit responds to only unmodelled spectral structure from faint sources. For clarity, no other calibration parameters are included. The observed error in the reflection calibration parameters is less than 1 part in 10$^3$. 

\begin{figure}
	\includegraphics[width = \columnwidth]{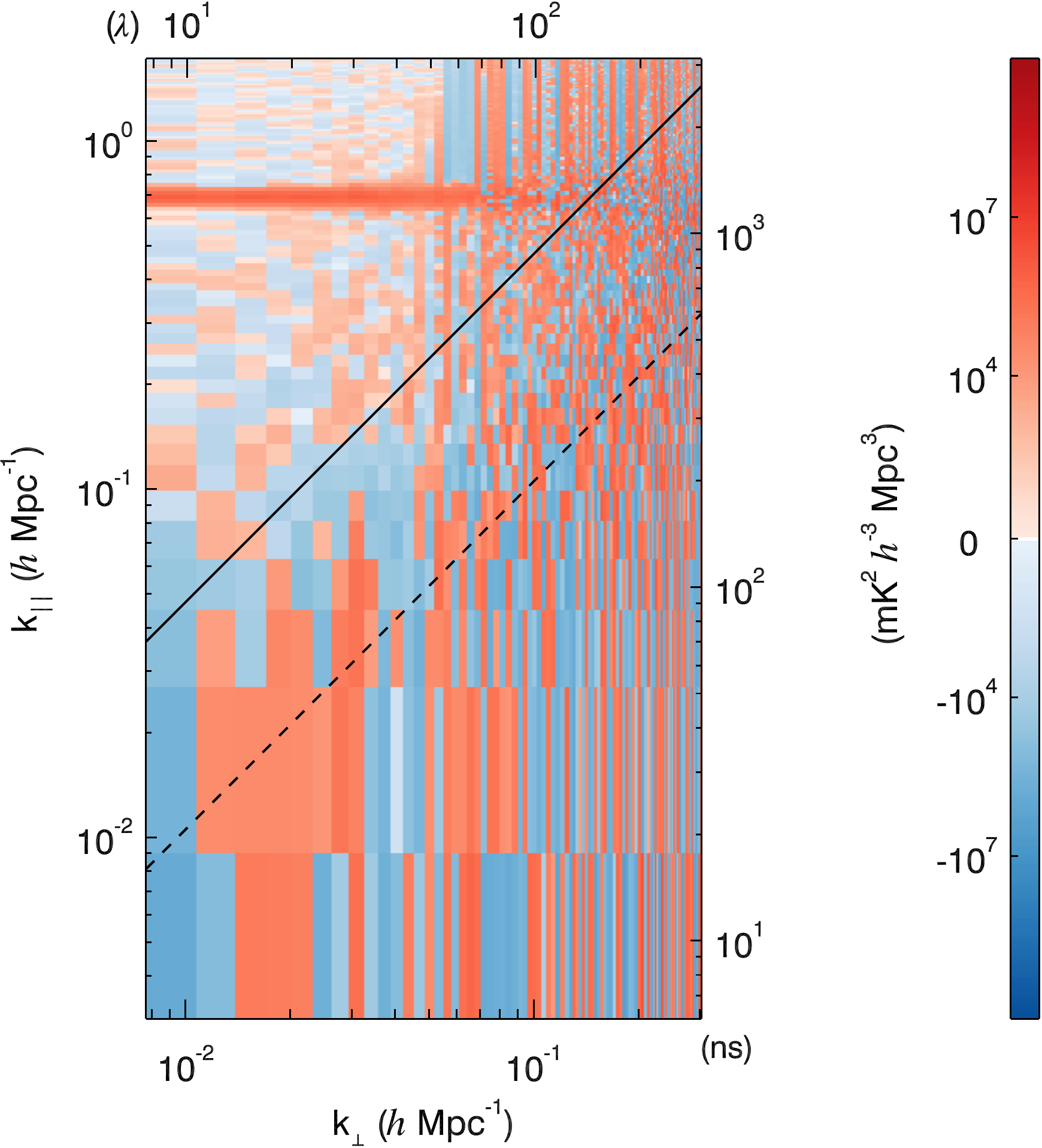}
	\caption{The difference 2D PS between the reference 2D PS and a 2D PS where a 150\,m cable reflection is fit to within 1 part in 10$^3$. Red indicates an excess of power, and blue indicates a depression of power. Spectral contamination occurs when there is error in the fit, and mode-mixing occurs between bright, intrinsic foregrounds and the mode associated with the fit error. Excess power at $k_\parallel\sim0.7$\,\emph{h}\,Mpc$^{-1}$ is over $10^{7}$\,mK$^2$\,\emph{h}$^{-3}$\,Mpc$^3$. This essentially removes that mode from EoR measurements.}
	\label{fig:150fit}
\end{figure}

The effect of coupling the intrinsic foregrounds and the imperfect fitted mode is shown in Figure~\ref{fig:150fit}. The difference 2D PS between the reference PS and the PS with a cable reflection calibration shows a clear excess of power at $k_\parallel\sim0.7$\,\emph{h}\,Mpc$^{-1}$, or the $k_\parallel$ associated with a 150\,m spectral ripple. 

The amount of excess power, about $10^{6}$\,mK$^2$\emph{h}$^{-3}$\,Mpc$^3$, is not a coincidence. The accuracy levels of the fit and the amplitude deviations in traditional per-frequency calibration in \S\ref{sec:trad} were similar. Whether the calibration is described per-frequency (\S\ref{sec:trad}) or per-spectral-mode ($\propto k_\parallel$), the same number of calibration terms are required to cover the bandwidth and the same level of contamination results.

Fitting for instrumental spectral structure necessarily removes those modes from potential measurement of the EoR. If the number of instrumental spectral features that must be calibrated fills much of the EoR window, measurement of the signal will be infeasible. This necessitates a smooth instrumental response in the modes associated with the EoR window.

As an example of how this affects instrument design, the MWA has chosen to only have a few different cable lengths in the array. This limits the regions of contamination within the the EoR window --- having many different cable lengths would contaminate the entire measurement region. This has also lead to a hard limit of all analog signal paths being <35~m in HERA.

%%%%%%%%%%%%%%%%%%%%%%%%%%%

%%%%%%%%%%%%%%%%%%%%%%%%%%%
\section{Mitigation by averaging calibration solutions}
\label{sec:bp}

Calibrating fine-scale instrumental frequency structure requires accurately modelling the antenna response while avoiding spurious spectral structure from unmodelled sources. This faint, unmodelled structure can be largely incoherent between antennas and can change quickly with Local Sidereal Time (LST) (due to an effect known as $uv$ rotation). We explore calibration techniques that average over antenna responses and over sidereal time to reduce the spectral contamination identified in the previous sections. 

Each antenna calibration solution is calculated from the subset of visibilities which include that antenna. The associated ``antenna PSF'' captures the effect of unmodelled sources on that specific calibration solution. Since each antenna's baseline coverage is largely independent for non-redundant arrays, the spurious spectral structure from unmodelled sources varies from antenna to antenna. If the antennae are identical in manufacture, then averaging their calibration solutions to form a common bandpass can reduce the calibration amplitude and phase deviations that cause spectral contamination. This may not be true for arrays with redundant layouts or layouts with strong symmetries, where the antenna PSFs can be very similar.

Additionally, if the antenna calibrations are very stable in time, subsequent calibration solutions can be averaged effectively. This relies on rotation of the antenna PSFs with LST to provide semi-independent contamination from unmodelled sources. 

We simulate a hypothetical array with the MWA layout (very random distribution, \citealt{beardsley_new_2012,beardsley_eor_2013,tingay_murchison_2013}) with mechanically identical and stable antennae. Using traditional calibration techniques, we calculate solutions per frequency channel for each of the 128 antennae in the MWA every 2 minutes for a 30 minute observation traversing zenith. The resulting 1920 solutions per frequency (128 x 15) are then averaged, excluding outliers beyond a $2\sigma$ cut. The final averaged per-frequency calibration solution is then applied to all antennae for only one observation, allowing direct comparison to the other simulations.

\begin{figure}
	\includegraphics[width = \columnwidth]{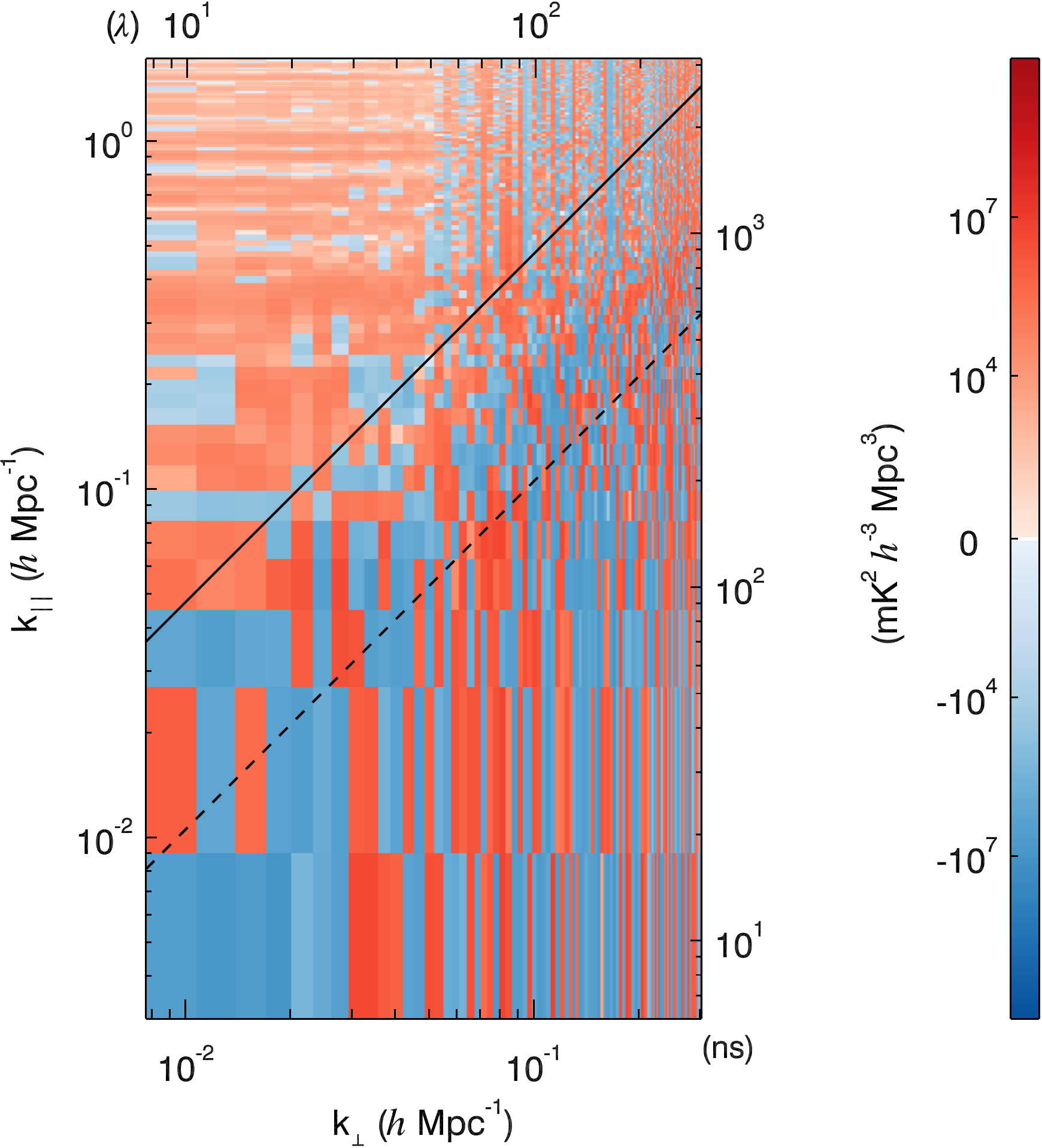}
	\caption{The difference 2D PS between a reference 2D PS and a 2D PS with an averaged calibration. Red indicates an excess of power, and blue indicates a depression of power. Calibration parameters per frequency were averaged across all antennae and over 30 minutes of observations for maximal LST coverage, excluding outliers beyond $2\sigma$. Excess power levels are much lower in the EoR window than with traditional calibration in Figure~\ref{fig:trad_diff}, but still higher than with smooth mitigation calibration techniques in Figure~\ref{fig:smooth}.}
	\label{fig:saved}
\end{figure}

Figure~\ref{fig:saved} presents the difference 2D PS between a reference PS and the averaged calibration solution PS. The amount of excess power in the EoR window has decreased by over three orders of magnitude compared to Figure~\ref{fig:trad_diff} with traditional per-frequency and per-antenna calibration. However, relative excess power still remains and lines of constant k$_\parallel$ contaminate the EoR window, indicating that spectral structure from unmodelled faint sources is still present in the average bandpass solution. The excess power level is similar to the expected power of the EoR. Whether or not an EoR measurement is feasible with averaged calibration will depend on the completeness of the sky model, similarity of the antennae across elements and time, and the instrument's design. 

\begin{figure*}
	\includegraphics[width = \textwidth]{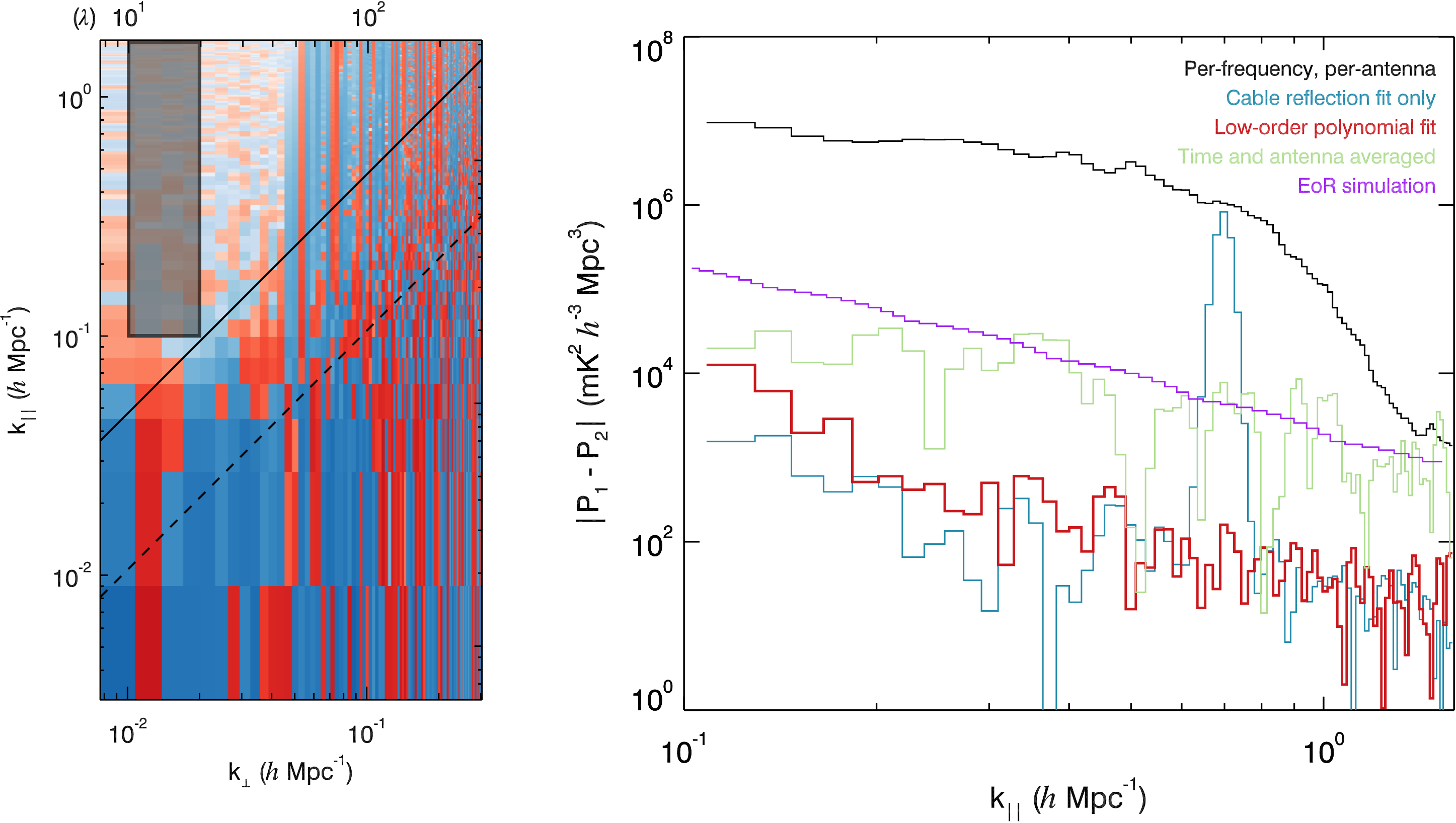}
	\caption{A 1D comparison between all PS differences using the various calibration techniques in this work (Figures~\ref{fig:trad_diff} -- \ref{fig:saved}) in the EoR window. We select the same k$_{||}$ and k$_{\perp}$ region for all PS difference plots, shown for illustration on a copy of Figure~\ref{fig:smooth} as a grey box. We then average k$_{\perp}$ within the box from 10 to 20\,$\lambda$ to generate 1D comparisons as a function of k$_{||}$. The plot on the right is the 1D comparison as a function of k$_{||}$ of the traditional per-frequency and per-antenna calibration (black, Fig.~\ref{fig:trad_diff}), the 150\,m cable reflection fit (blue, Fig.~\ref{fig:150fit}), the low-order polynomial fit calibration (red, Fig.~\ref{fig:smooth}), and the maximally-averaged LST and antenna element calibration (green, Fig.~\ref{fig:saved}). The PS of the estimated EoR (purple, Fig.~\ref{fig:eor}) is also plotted to show where contamination will surpass the desired detection. A maximally-averaged calibration at the current level of precision for the MWA is not a practical solution given that this work only explores one of many possible contamination sources, and contamination is already more or less at the level of the EoR. The low-order polynomial fit is by far the best solution if the instrument varies smoothly across the entire frequency band.}
	\label{fig:1d}
\end{figure*}

Figure~\ref{fig:1d} compares all difference PS in this work to the expected level of the EoR. We average k$_{\perp}$ from 10 to 20\,$\lambda$ to generate a 1D PS as a function of k$_{||}$ in the EoR window. The level of contamination should be significantly below the EoR in order to realistically detect it with all other possible sources of error not explored in this work. We find that the contamination from the maximally-averaged calibration solution over all possible LSTs and antennae using a 4000 source sky model is at the level of the EoR, and therefore not a practical solution for the MWA. In contrast, we find that using low-order polynomials described in \S\ref{sec:smooth_full} is the best calibration method; it is lower than the expected EoR by one to two orders of magnitude. Current efforts to calibrate MWA EoR data use antenna and time averaging in conjunction with the smooth characteristics of the antenna to reduce the level of calibration contamination \citep{beardsley_thesis}. Other instruments may be able to achieve practical levels of spectral contamination with only averaged calibration solutions if more LST or antenna samples can be used.

In practice, thermal noise will also effect the PS and the calibration solutions. An additional set of simulations explored the effect of thermal noise on the calibration solutions, and showed the expected additional spectral contamination in the EoR window. The thermal noise contribution is uncorrelated in time, and averaging calibration solutions night to night proves effective in removing this contribution (see \citealt{trott_2016} for detailed analysis). However, spectral contamination from faint, unmodelled sources still remains as a systematic error due to the limited number of antennae and observing LSTs. The systematic contribution of the calibration due to an imperfect calibration catalogue has more potential than thermal noise to hinder long integrations in search for the EoR.
	
%%%%%%%%%%%%%%%%%%%%%%%%%%%

%%%%%%%%%%%%%%%%%%%%%%%%%%
\section{Discussion}
\label{sec:conclusions}

This work explores the impact of instrumental calibration on EoR PS measurements using a precision end-to-end simulation. Our simulation framework includes a precise frequency-dependent instrument model, a foreground model based on the observed compact sources in the MWA EoR0 field, and the full FHD and \eppsilon{} calibration, imaging, and PS estimation pipeline used in EoR analysis \citep{beardsley_thesis,jacobs_murchison_2016}. We found that it was crucial to include a calibration catalogue that does not exactly match the simulated sky and to propagate all of the calibration effects to the PS where the EoR measurement will be performed. 

In \S\ref{sec:methods}, we introduced the simulation pipeline and in Figure~\ref{fig:eor} showed that the EoR is perfectly recovered if the calibration model is identical to the foreground sources in the sky model. However, it is impossible to have a perfect calibration model --- any catalogue will miss faint sources and have small errors in the flux, position, and spatial morphology of the included sources. In \S\ref{sec:trad}, we simulated the effect of an imperfect calibration catalogue by using only the brightest 4,000 sources for calibrating and a deeper 6,950 sources in the foreground sky simulation. Using a traditional per-frequency antenna calibration, we showed that the resulting power spectrum has contamination throughout the EoR window --- precluding EoR observations.

\S\ref{sec:trad} explored the source of this contamination, identifying the PSFs of the faint, unmodelled sources as the root cause. The chromatic PSFs of the unmodelled sources lead to small calibration errors on the order of $10^{-3}$ that couple to the bright foregrounds and distribute spectral contamination throughout the EoR window (Figure~\ref{fig:trad_diff}). \S\ref{sec:smooth_full} \& \S\ref{sec:bp} then explored more advanced calibration techniques that could mitigate this contamination. 

The lesson for SKA, HERA, the MWA upgrade, the LOFAR upgrade, and other future EoR instruments is that any spectral features of the antennae that are calibrated will lead to contamination at the corresponding location in the EoR window. This is most clearly seen in Figure~\ref{fig:150fit}. This source of contamination via calibration errors places strong constraints on the instruments and planned observational programmes. To avoid the contamination identified in this paper, we identify four potential solutions:

\begin{enumerate}
\vspace{-5pt}
\renewcommand{\theenumi}{(\arabic{enumi})}
    \item Create a nearly perfect calibration catalogue. As the quality of the calibration catalogue improves, the amplitude of the associated calibration errors and modulated PS contamination both decrease. The necessary precision of the catalogue depends on the details of the array (in particular the PSF over the calibration period), and can be simulated using the techniques developed in \S\ref{sec:methods}. This solution requires that a high fidelity foreground catalogue be created before EoR analysis can start, and for some arrays creating a catalogue to the necessary precision may not be possible.
    \item Use antennae with very smooth spectral responses. It is possible to design an analog and digital system that is naturally very spectrally smooth, with no calibration of features within the EoR window needed. The PS of such a hypothetical antenna is shown in Figure~\ref{fig:trad_diff}. In practice, this means there can be no spectral features larger than $\sim$10$^{-5}$ in the antenna or receiver system with spectral scales faster than $\sim$8~MHz (125\,ns). This thinking is driving the spectrally smooth antenna and receiver designs of HERA and the MWA upgrade \citep{neben_hydrogen_2016, ewall-wice_hera_2016, thyagarajan_effects_2016}. %Patra et. al. in prep
    \item Manufacture physically identical and stable antennae. If all the antennae have the same spectral features and their response is very stable in time, then the antenna-time averaging explored in \S\ref{sec:bp} can be used to reduce the spectral smoothness and catalogue precision requirements. While this work focuses on sky-based calibration, redundant calibration techniques explicitly depend on the antenna responses being identical \citep{wieringa_investigation_1992,liu_precision_2010,parsons_sensitivity_2012,noorishad_redundancy_2012,zheng_miteor:_2014}.
    \item Develop an external calibration system. The coupling to an incomplete sky catalogue can be entirely avoided by using an external calibrator such as a drone, satellite, pulsar, or pulse injection system. It is still a challenge to reach the $\sim$10$^{-5}$ calibration precision needed, but several groups have been pursuing this path \citep{ newburgh_calibrating_2014,patra_bandpass_2015,neben_measuring_2015}.
\end{enumerate}
Which combinations of these four approaches will work the best is jointly dependent on the instrument-specific antenna PSFs and the precision and depth of the calibration catalogue. Calibration simulations following the techniques developed in \S\ref{sec:methods} must be explored for each instrument to calculate the necessary instrument specifications.

In our opinion, building antennae with a naturally smooth spectral response (2) is the lowest risk and most cost-effective approach. Basing analysis plans on the development of a nearly perfect calibration catalog (1) is risky because it is hard to predict the achievable precision and depth of a catalogue made with a new instrument. Manufacturing physically identical antennae (3) is expensive, particularly factoring in the logistics of maintaining identical performance in the field. Similarly, developing external calibration systems of the requisite precision (4) is expected to be expensive. 

Spectrally smooth antennae described in approach (2) are the practical solution to avoiding spectral contamination of the EoR PS window. Figure~\ref{fig:1d} shows that it is the least contaminated calibration method in sensitive EoR modes explored in this work. We recommend that EoR instruments aim to have no spectral features larger than $\sim$10$^{-5}$ on scales faster than $\sim$8~MHz (125\,ns).

It has long been recognized that precision calibration is necessary to perform EoR PS observations. In this work, we have identified that traditional per-frequency calibration techniques with an imperfect calibration catalogue can lead to significant contamination of the EoR PS window. We feel this insight and the associated simulation techniques can help guide the design of the SKA and other future EoR machines.

%%%%%%%%%%%%%%%%%%%%%%%%%%%

\section*{Acknowledgements}

We would like to thank Cathryn Trott and Aaron Ewall-Wice for their insightful remarks and guidance. This work was supported by National Science Foundation grants AST-1410484 and AST-1506024.

%%%%%%%%%%%%%%%%%%%%%%%%%%%%%%%%%%%%%%%%%%%%%%%%%%


%\clearpage

\bibliographystyle{mnras}
\bibliography{cal_sim} % if your bibtex file is called example.bib

\newcommand{\SortNoop}[1]{}
\begin{thebibliography}{}
\makeatletter
\relax
\def\mn@urlcharsother{\let\do\@makeother \do\$\do\&\do\#\do\^\do\_\do\%\do\~}
\def\mn@doi{\begingroup\mn@urlcharsother \@ifnextchar [ {\mn@doi@}
  {\mn@doi@[]}}
\def\mn@doi@[#1]#2{\def\@tempa{#1}\ifx\@tempa\@empty \href
  {http://dx.doi.org/#2} {doi:#2}\else \href {http://dx.doi.org/#2} {#1}\fi
  \endgroup}
\def\mn@eprint#1#2{\mn@eprint@#1:#2::\@nil}
\def\mn@eprint@arXiv#1{\href {http://arxiv.org/abs/#1} {{\tt arXiv:#1}}}
\def\mn@eprint@dblp#1{\href {http://dblp.uni-trier.de/rec/bibtex/#1.xml}
  {dblp:#1}}
\def\mn@eprint@#1:#2:#3:#4\@nil{\def\@tempa {#1}\def\@tempb {#2}\def\@tempc
  {#3}\ifx \@tempc \@empty \let \@tempc \@tempb \let \@tempb \@tempa \fi \ifx
  \@tempb \@empty \def\@tempb {arXiv}\fi \@ifundefined
  {mn@eprint@\@tempb}{\@tempb:\@tempc}{\expandafter \expandafter \csname
  mn@eprint@\@tempb\endcsname \expandafter{\@tempc}}}

\bibitem[\protect\citeauthoryear{Beardsley}{Beardsley}{2015}]{beardsley_thesis}
Beardsley A.,  2015, PhD thesis, University of Washington

\bibitem[\protect\citeauthoryear{Beardsley et~al.,}{Beardsley
  et~al.}{2012}]{beardsley_new_2012}
Beardsley A.~P.,  et~al., 2012, \mn@doi [Monthly Notices of the Royal
  Astronomical Society] {10.1111/j.1365-2966.2012.20878.x}, 425, 1781

\bibitem[\protect\citeauthoryear{Beardsley et~al.,}{Beardsley
  et~al.}{2013}]{beardsley_eor_2013}
Beardsley A.~P.,  et~al., 2013, \mn@doi [Monthly Notices of the Royal
  Astronomical Society: Letters] {10.1093/mnrasl/sls013}, 429, L5

\bibitem[\protect\citeauthoryear{Bowman et~al.,}{Bowman
  et~al.}{2013}]{bowman_science_2013}
Bowman J.~D.,  et~al., 2013, \mn@doi [PASA - Publications of the Astronomical
  Society of Australia] {10.1017/pas.2013.009}, 30

\bibitem[\protect\citeauthoryear{Datta, Bhatnagar  \& Carilli}{Datta
  et~al.}{2009}]{datta_detection_2009}
Datta A.,  Bhatnagar S.,   Carilli C.~L.,  2009, \mn@doi [The Astrophysical
  Journal] {10.1088/0004-637X/703/2/1851}, 703, 1851

\bibitem[\protect\citeauthoryear{Datta, Bowman  \& Carilli}{Datta
  et~al.}{2010}]{datta_bright_2010}
Datta A.,  Bowman J.~D.,   Carilli C.~L.,  2010, \mn@doi [The Astrophysical
  Journal] {10.1088/0004-637X/724/1/526}, 724, 526

\bibitem[\protect\citeauthoryear{Dillon et~al.,}{Dillon
  et~al.}{2015}]{dillon_empirical_2015}
Dillon J.~S.,  et~al., 2015, \mn@doi [Physical Review D]
  {10.1103/PhysRevD.91.123011}, 91, 123011

\bibitem[\protect\citeauthoryear{{Ewall-Wice} et~al.,}{{Ewall-Wice}
  et~al.}{2016a}]{ewall-wice_2016_xray}
{Ewall-Wice} A.,  et~al., 2016a, Monthly Notices of the Royal Astronomical
  Society

\bibitem[\protect\citeauthoryear{Ewall-Wice et~al.,}{Ewall-Wice
  et~al.}{2016b}]{ewall-wice_hera_2016}
Ewall-Wice A.,  et~al., 2016b, The {HERA} {Dish} {II}: {Electromagnetic}
  {Simulations} and {Science} {Implications} (\mn@eprint {arXiv} {1602.06277})

\bibitem[\protect\citeauthoryear{Fomalont \& Perley}{Fomalont \&
  Perley}{1999}]{fomalont_1999}
Fomalont E.~B.,  Perley R.~A.,  1999, in Taylor G.~B.,  Carilli C.~L.,   Perley
  R.~A.,  eds, Synthesis Imaging in Radio Astronomy II. ASP Conference series,
  pp 79--110

\bibitem[\protect\citeauthoryear{Furlanetto, Peng~Oh  \& Briggs}{Furlanetto
  et~al.}{2006}]{furlanetto_cosmology_2006}
Furlanetto S.~R.,  Peng~Oh S.,   Briggs F.~H.,  2006, \mn@doi [Physics Reports]
  {10.1016/j.physrep.2006.08.002}, 433, 181

\bibitem[\protect\citeauthoryear{{\SortNoop{Haarlem}}van~Haarlem
  et~al.,}{{\SortNoop{Haarlem}}van~Haarlem
  et~al.}{2013}]{van_haarlem_lofar:_2013}
{\SortNoop{Haarlem}}van~Haarlem M.~P.,  et~al., 2013, \mn@doi [Astronomy \&
  Astrophysics] {10.1051/0004-6361/201220873}, 556, A2

\bibitem[\protect\citeauthoryear{Hazelton, Morales  \& Sullivan}{Hazelton
  et~al.}{2013}]{hazelton_fundamental_2013}
Hazelton B.~J.,  Morales M.~F.,   Sullivan I.~S.,  2013, \mn@doi [Astrophysical
  Journal] {10.1088/0004-637X/770/2/156}, 770, 156

\bibitem[\protect\citeauthoryear{Hogg}{Hogg}{1999}]{hogg_distance_1999}
Hogg D.~W.,  1999, Distance measures in cosmology, arXiv: astro-ph/9905116,
  \url {http://arxiv.org/abs/astro-ph/9905116}

\bibitem[\protect\citeauthoryear{Jacobs et~al.,}{Jacobs
  et~al.}{2016}]{jacobs_murchison_2016}
Jacobs D.~C.,  et~al., 2016, The Astrophysical Journal

\bibitem[\protect\citeauthoryear{Koopmans et~al.,}{Koopmans
  et~al.}{2015}]{koopmans_cosmic_2015}
Koopmans L.,  et~al., 2015, in Bourke T.~L.,  et~al., eds, Advancing
  Astrophysics with the Square Kilometre Array (AASKA14). SISSA

\bibitem[\protect\citeauthoryear{Liu, Tegmark, Morrison, Lutomirski  \&
  Zaldarriaga}{Liu et~al.}{2010}]{liu_precision_2010}
Liu A.,  Tegmark M.,  Morrison S.,  Lutomirski A.,   Zaldarriaga M.,  2010,
  \mn@doi [Monthly Notices of the Royal Astronomical Society]
  {10.1111/j.1365-2966.2010.17174.x}, 408, 1029

\bibitem[\protect\citeauthoryear{Liu, Parsons  \& Trott}{Liu
  et~al.}{2014}]{liu_epoch_2014}
Liu A.,  Parsons A.~R.,   Trott C.~M.,  2014, \mn@doi [Physical Review D]
  {10.1103/PhysRevD.90.023018}, 90, 023018

\bibitem[\protect\citeauthoryear{Lonsdale et~al.,}{Lonsdale
  et~al.}{2009}]{lonsdale_murchison_2009}
Lonsdale C.,  et~al., 2009, \mn@doi [Proceedings of the IEEE]
  {10.1109/JPROC.2009.2017564}, 97, 1497

\bibitem[\protect\citeauthoryear{Matteo, Perna, Abel  \& Rees}{Matteo
  et~al.}{2002}]{matteo_radio_2002}
Matteo T.~D.,  Perna R.,  Abel T.,   Rees M.~J.,  2002, \mn@doi [The
  Astrophysical Journal] {10.1086/324293}, 564, 576

\bibitem[\protect\citeauthoryear{Mellema et~al.,}{Mellema
  et~al.}{2013}]{mellema_reionization_2013}
Mellema G.,  et~al., 2013, \mn@doi [Experimental Astronomy]
  {10.1007/s10686-013-9334-5}, 36, 235

\bibitem[\protect\citeauthoryear{Mitchell, Greenhill, Wayth, Sault, Lonsdale,
  Cappallo, Morales  \& Ord}{Mitchell et~al.}{2008}]{mitchell_real-time_2008}
Mitchell D.,  Greenhill L.,  Wayth R.,  Sault R.,  Lonsdale C.,  Cappallo R.,
  Morales M.,   Ord S.,  2008, \mn@doi [IEEE Journal of Selected Topics in
  Signal Processing] {10.1109/JSTSP.2008.2005327}, 2, 707

\bibitem[\protect\citeauthoryear{Morales \& Hewitt}{Morales \&
  Hewitt}{2004}]{morales_toward_2004}
Morales M.~F.,  Hewitt J.,  2004, \mn@doi [The Astrophysical Journal]
  {10.1086/424437}, 615, 7

\bibitem[\protect\citeauthoryear{Morales \& Wyithe}{Morales \&
  Wyithe}{2010}]{morales_reionization_2010}
Morales M.~F.,  Wyithe J. S.~B.,  2010, \mn@doi [Annual Review of Astronomy and
  Astrophysics] {10.1146/annurev-astro-081309-130936}, 48, 127

\bibitem[\protect\citeauthoryear{Morales, Hazelton, Sullivan  \&
  Beardsley}{Morales et~al.}{2012}]{morales_four_2012}
Morales M.~F.,  Hazelton B.,  Sullivan I.,   Beardsley A.,  2012, \mn@doi [The
  Astrophysical Journal] {10.1088/0004-637X/752/2/137}, 752, 137

\bibitem[\protect\citeauthoryear{Neben et~al.,}{Neben
  et~al.}{2015}]{neben_measuring_2015}
Neben A.~R.,  et~al., 2015, \mn@doi [Radio Science] {10.1002/2015RS005678}, 50,
  2015RS005678

\bibitem[\protect\citeauthoryear{Neben et~al.,}{Neben
  et~al.}{2016}]{neben_hydrogen_2016}
Neben A.~R.,  et~al., 2016, The {Hydrogen} {Epoch} of {Reionization} {Array}
  {Dish} {I}: {Beam} {Pattern} {Measurements} and {Science} {Implications}
  (\mn@eprint {arXiv} {1602.03887})

\bibitem[\protect\citeauthoryear{Newburgh et~al.,}{Newburgh
  et~al.}{2014}]{newburgh_calibrating_2014}
Newburgh L.~B.,  et~al., 2014, in Ground-based and Airborne Telescopes V. Proc.
  SPIE 9145, pp 91454V--91454V--18, \mn@doi{10.1117/12.2056962}

\bibitem[\protect\citeauthoryear{Ng \& See}{Ng \&
  See}{1996}]{ng_sensor-array_1996}
Ng B.~C.,  See C. M.~S.,  1996, \mn@doi [IEEE Transactions on Antennas and
  Propagation] {10.1109/8.509886}, 44, 827

\bibitem[\protect\citeauthoryear{Noorishad, Wijnholds, van Ardenne  \& van~der
  Hulst}{Noorishad et~al.}{2012}]{noorishad_redundancy_2012}
Noorishad P.,  Wijnholds S.~J.,  van Ardenne A.,   van~der Hulst J.~M.,  2012,
  \mn@doi [Astronomy \& Astrophysics] {10.1051/0004-6361/201219087}, 545, A108

\bibitem[\protect\citeauthoryear{Offringa et~al.,}{Offringa
  et~al.}{2016}]{offringa_parametrising_2016}
Offringa A.~R.,  et~al., 2016, \mn@doi [Monthly Notices of the Royal
  Astronomical Society] {10.1093/mnras/stw310}, p. stw310

\bibitem[\protect\citeauthoryear{Parsons et~al.,}{Parsons
  et~al.}{2010}]{parsons_precision_2010}
Parsons A.~R.,  et~al., 2010, \mn@doi [The Astronomical Journal]
  {10.1088/0004-6256/139/4/1468}, 139, 1468

\bibitem[\protect\citeauthoryear{Parsons, Pober, McQuinn, Jacobs  \&
  Aguirre}{Parsons et~al.}{2012a}]{parsons_sensitivity_2012}
Parsons A.,  Pober J.,  McQuinn M.,  Jacobs D.,   Aguirre J.,  2012a, \mn@doi
  [The Astrophysical Journal] {10.1088/0004-637X/753/1/81}, 753, 81

\bibitem[\protect\citeauthoryear{Parsons, Pober, Aguirre, Carilli, Jacobs  \&
  Moore}{Parsons et~al.}{2012b}]{parsons_per-baseline_2012}
Parsons A.~R.,  Pober J.~C.,  Aguirre J.~E.,  Carilli C.~L.,  Jacobs D.~C.,
  Moore D.~F.,  2012b, \mn@doi [The Astrophysical Journal]
  {10.1088/0004-637X/756/2/165}, 756, 165

\bibitem[\protect\citeauthoryear{Patra, Bray, Ekers  \& Roberts}{Patra
  et~al.}{2015}]{patra_bandpass_2015}
Patra N.,  Bray J.,  Ekers R.,   Roberts P.,  2015, arXiv:1502.05862 [astro-ph]

\bibitem[\protect\citeauthoryear{Pearson \& Readhead}{Pearson \&
  Readhead}{1984}]{pearson_image_1984}
Pearson T.~J.,  Readhead A. C.~S.,  1984, \mn@doi [Annual Review of Astronomy
  and Astrophysics] {10.1146/annurev.aa.22.090184.000525}, 22, 97

\bibitem[\protect\citeauthoryear{Peng~Oh \& Mack}{Peng~Oh \&
  Mack}{2003}]{peng_oh_foregrounds_2003}
Peng~Oh S.,  Mack K.~J.,  2003, \mn@doi [Monthly Notices of the Royal
  Astronomical Society] {10.1111/j.1365-2966.2003.07133.x}, 346, 871

\bibitem[\protect\citeauthoryear{Pober et~al.,}{Pober
  et~al.}{2013}]{pober_opening_2013}
Pober J.~C.,  et~al., 2013, \mn@doi [The Astrophysical Journal Letters]
  {10.1088/2041-8205/768/2/L36}, 768, L36

\bibitem[\protect\citeauthoryear{Pober et~al.,}{Pober
  et~al.}{2014}]{pober_what_2014}
Pober J.~C.,  et~al., 2014, \mn@doi [The Astrophysical Journal]
  {10.1088/0004-637X/782/2/66}, 782, 66

\bibitem[\protect\citeauthoryear{Pober et~al.,}{Pober
  et~al.}{2016}]{pober_importance_2016}
Pober J.~C.,  et~al., 2016, \mn@doi [The Astrophysical Journal]
  {10.3847/0004-637X/819/1/8}, 819, 8

\bibitem[\protect\citeauthoryear{Salvini \& Wijnholds}{Salvini \&
  Wijnholds}{2014}]{salvini_fast_2014}
Salvini S.,  Wijnholds S.~J.,  2014, \mn@doi [Astronomy \& Astrophysics]
  {10.1051/0004-6361/201424487}, 571, A97

\bibitem[\protect\citeauthoryear{Sullivan}{Sullivan}{2009}]{iii_cosmic_2009}
Sullivan III W.~T.,  2009, Cosmic {Noise}: {A} {History} of {Early} {Radio}
  {Astronomy}.
Cambridge University Press

\bibitem[\protect\citeauthoryear{Sullivan et~al.,}{Sullivan
  et~al.}{2012}]{sullivan_fast_2012}
Sullivan I.~S.,  et~al., 2012, \mn@doi [The Astrophysical Journal]
  {10.1088/0004-637X/759/1/17}, 759, 17

\bibitem[\protect\citeauthoryear{Switzer, Chang, Masui, Pen  \& Voytek}{Switzer
  et~al.}{2015}]{switzer_interpreting_2015}
Switzer E.~R.,  Chang T.-C.,  Masui K.~W.,  Pen U.-L.,   Voytek T.~C.,  2015,
  \mn@doi [Astrophysical Journal] {10.1088/0004-637X/815/1/51}, 815, 51

\bibitem[\protect\citeauthoryear{Thyagarajan et~al.,}{Thyagarajan
  et~al.}{2013}]{thyagarajan_study_2013}
Thyagarajan N.,  et~al., 2013, \mn@doi [The Astrophysical Journal]
  {10.1088/0004-637X/776/1/6}, 776, 6

\bibitem[\protect\citeauthoryear{Thyagarajan, Parsons, DeBoer, Bowman,
  Ewall-Wice, Neben  \& Patra}{Thyagarajan
  et~al.}{2016}]{thyagarajan_effects_2016}
Thyagarajan N.,  Parsons A.,  DeBoer D.,  Bowman J.,  Ewall-Wice A.,  Neben A.,
    Patra N.,  2016, Effects of {Antenna} {Beam} {Chromaticity} on {Redshifted}
  21{\textasciitilde}cm {Power} {Spectrum} and {Implications} for {Hydrogen}
  {Epoch} of {Reionization} {Array} (\mn@eprint {arXiv} {1603.08958})

\bibitem[\protect\citeauthoryear{Tingay et~al.,}{Tingay
  et~al.}{2013}]{tingay_murchison_2013}
Tingay S.~J.,  et~al., 2013, \mn@doi [PASA - Publications of the Astronomical
  Society of Australia] {10.1017/pasa.2012.007}, 30

\bibitem[\protect\citeauthoryear{{\SortNoop{Tol}}van~der Tol, Jeffs  \& van~der
  Veen}{{\SortNoop{Tol}}van~der Tol
  et~al.}{2007}]{van_der_tol_self-calibration_2007}
{\SortNoop{Tol}}van~der Tol S.,  Jeffs B.,   van~der Veen A.,  2007, \mn@doi
  [IEEE Transactions on Signal Processing] {10.1109/TSP.2007.896243}, 55, 4497

\bibitem[\protect\citeauthoryear{Trott \& Wayth}{Trott \&
  Wayth}{2016}]{trott_2016}
Trott C.~M.,  Wayth R.~B.,  2016, \mn@doi [Publications of the Astronomical
  Society of Australia] {10.1017/pasa.2016.18}, 33

\bibitem[\protect\citeauthoryear{Trott, Wayth  \& Tingay}{Trott
  et~al.}{2012}]{trott_impact_2012}
Trott C.~M.,  Wayth R.~B.,   Tingay S.~J.,  2012, \mn@doi [The Astrophysical
  Journal] {10.1088/0004-637X/757/1/101}, 757, 101

\bibitem[\protect\citeauthoryear{Vedantham, Shankar  \& Subrahmanyan}{Vedantham
  et~al.}{2012}]{vedantham_imaging_2012}
Vedantham H.,  Shankar N.~U.,   Subrahmanyan R.,  2012, \mn@doi [The
  Astrophysical Journal] {10.1088/0004-637X/745/2/176}, 745, 176

\bibitem[\protect\citeauthoryear{Wieringa}{Wieringa}{1992}]{wieringa_investigation_1992}
Wieringa M.~H.,  1992, \mn@doi [Experimental Astronomy] {10.1007/BF00420576},
  2, 203

\bibitem[\protect\citeauthoryear{Wijnholds \& van~der Veen}{Wijnholds \&
  van~der Veen}{2009}]{wijnholds_van_der_veen_multisource_2009}
Wijnholds S.~J.,  van~der Veen A.-J.,  2009, \mn@doi [IEEE Transactions on
  Signal Processing] {10.1109/TSP.2009.2022894,
  info:doi/10.1109/TSP.2009.2022894}, 57, 3512

\bibitem[\protect\citeauthoryear{Yatawatta et~al.,}{Yatawatta
  et~al.}{2013}]{yatawatta_initial_2013}
Yatawatta S.,  et~al., 2013, \mn@doi [Astronomy \& Astrophysics]
  {10.1051/0004-6361/201220874}, 550, A136

\bibitem[\protect\citeauthoryear{Zheng et~al.,}{Zheng
  et~al.}{2014}]{zheng_miteor:_2014}
Zheng H.,  et~al., 2014, \mn@doi [Monthly Notices of the Royal Astronomical
  Society] {10.1093/mnras/stu1773}, 445, 1084

\makeatother
\end{thebibliography}

% Don't change these lines
\bsp	% typesetting comment
\label{lastpage}
\end{document}